\documentclass[%
 aip,
 amsmath,amssymb,%nobibnotes,
 preprint,%
]{revtex4}

\usepackage{graphicx}% Include figure files
\usepackage{dcolumn}% Align table columns on decimal point
\usepackage{bm}% bold math
\usepackage[utf8]{inputenc}
\usepackage[T1]{fontenc}
\usepackage{mathptmx} 

\begin{document}

\title{Experimental Search for Non-Newtonian Forces in the Nanometer Scale with Slow Neutrons}% Force line breaks with \\

\author{Y. Kamiya} % Write as First name Surname
 \email[Corresponding author: ]{kamiya@icepp.s.u-tokyo.ac.jp}
\affiliation{%
  International Center for Elementary Particle Physics and Graduate School of Science, the University of Tokyo,
7-3-1 Hongo, Bunkyo-ku, Tokyo 113-0033, Japan
}%
\author{R. Cubitt}
\affiliation{%
Institut Laue-Langevin, 71 avenue des Martyrs, CS 20156, 38042 Grenoble cedex 9, France
}%
\author{L. Porcar}
\affiliation{%
Institut Laue-Langevin, 71 avenue des Martyrs, CS 20156, 38042 Grenoble cedex 9, France
}%
\author{O. Zimmer}
\affiliation{%
Institut Laue-Langevin, 71 avenue des Martyrs, CS 20156, 38042 Grenoble cedex 9, France
}%

\author{G. N. Kim}
\affiliation{%
Department of Physics, Kyungpook National University, Daegu 702-701, Republic of Korea
}%
\author{S. Komamiya}
\affiliation{%
  International Center for Elementary Particle Physics and Graduate School of Science, the University of Tokyo,
7-3-1 Hongo, Bunkyo-ku, Tokyo 113-0033, Japan
}%

\date{\today} % It is always \today, today, but any date may be explicitly specified
              % Not printed for conference proceedings

\begin{abstract}
Improved limits for new gravity-like short-range interactions,
in which a scattering potential is modeled by the Yukawa-type parametrization,
have been obtained by measuring the angular distribution of 6 \AA\ neutrons scattering from atomic xenon gas.
We have collected  approximately $1.4 \times 10^8$ small-angle neutron scattering events.
The data are interpreted as no evidence of new forces and show improved upper limits 
on the coupling strength in the interaction range of $0.3$ nm to $9$ nm.
These improved constraints are also interpreted as new limits for a model, in which 
a charge of the new forces is expressed as a linear combination of the baryon number and the lepton number.
\end{abstract}

\maketitle

\section{Introduction}
The standard model of particle physics has been developed for almost half a century
and provides a solid basis for understanding nature from the view point of its fundamental constituents, matter and interactions.
It has succeeded in describing various phenomena of particle physics observed 
in a wide energy range up to the electroweak scale.
However, there are still some unnatural points left, 
such as the so-called fine-tuning problem 
in radiative corrections to the Higgs mass, 
or the metastability of the electroweak vacuum
\cite{Espinosa:1995, Isidori:2001, Giudice:2013}.
Those imply extensions to the standard model or new theoretical frameworks 
which may appear at an energy below the Grand Unified Theory's (GUT's) scale \cite{Georgi:1974sy}.

\section{Experimental Scheme}
We  focus particularly on searching for new 
bosons of gravity-like coupling fields,
which are naturally involved in some beyond-the-standard-model theories 
based on supersymmetry or spatial extra dimensions 
\cite{Fayet:1993fu, ArkaniHamed:1998gb, Randall:389728}.
They are also predicted by scalar-tensor theories of gravitation that treat gravity as a field theory 
\cite{Brans:1961fv, Khoury:2004kv}.
In many cases, new fields would become manifest in additional gravity-like forces between test objects.

A sensitive method to search for these effects evaluates the neutron scattering processes with xenon atomic gas.
The experiment has been performed using a monochromatic cold neutron beam 
on the D22  small-angle neutron scattering instrument at the Institut Laue-Langevin (ILL) \cite{ILLProposal},
and a momentum transfer dependence ($q$-dependence) in the scattering processes has been analyzed.
We have selected a slightly larger neutron wavelength than in a previous measurement \cite{Kamiya:2015go},
allowing us to investigate a region of lower $q$.
In addition, we have collected around $1.4 \times 10^8$ scattering events.
Along with this improvement of statistical error, 
systematic control on the xenon gas pressure has been carefully considered.
We follow the evaluation scheme which was widely utilized in previously performed experiments
\cite{Pokotilovski:2006gd, Haddock:2018hf, Kamiya:2015go, Nesvizhevsky:2008jz, Borkowski:2019, BORDAG:2001tt}.
Therein, the Yukawa-type scattering potential in natural units,
\begin{equation}
V_{new}(r) = - \frac{1}{4\pi} g^2 Q_1 Q_2 \frac{e^{-\mu r}}{r}~,
\end{equation}
is used as a model of new physics,
where $g^2$ is a coupling strength, $Q_i$ are coupling charges,
and $\mu$ is the mass of the boson mediating the new interaction.
This additional potential causes a new $q$-dependent term in neutron scattering distributions.
The differential scattering cross-section can in total be written as
\begin{equation}
\frac{d\sigma}{d\Omega} (q) \simeq  b_c^2 \left\{ 1 + 2\chi_{em}[1-f(q)] + 2\chi_{new} \frac{\mu^2}{q^2 + \mu^2} \right\}~, \label{eq:cs}
\end{equation}
where the constant $b_c$ ($\sim 5$ fm) is the coherent scattering length.
The dominating first term describes ordinary nuclear scattering.
The second term represents the well-known $q$-dependent electromagnetic interactions 
between the neutron and intra-atomic fields.
Here, $\chi_{em}$ is on the order of $10^{-2}$.
$f(q)$ is the atomic form factor, which is modeled as 
$f(q) = [1+3(q/q_0)^2]^{-0.5}$ with $q_0 = 6.86$ \AA$^{-1}$ for xenon gas within $ 10^{-4}$ accuracy \cite{Sears:1986go}.
The third term is due to the additional $q$-dependent Yukawa-type scattering potential.
Here $\chi_{new} \equiv m_n g^2 Q_1 Q_2 /2\pi b_c \mu^2$ where $m_n$ is the neutron mass.
Since $\chi_{em}$ and $\chi_{new}$ are small enough, we neglect terms where products of the small amplitudes occur.
We assume that the coupling charges are masses $M_i$ to keep compatibility 
with discussions on the Cavendish type experiments \cite{BORDAG:2001tt, Chen:2016ch, Geraci:2008hp, Kapner:2007hh, Tan:2016ix, Yang:2012dy}, 
which verify the inverse-square law of Newtonian gravity using macroscopic test masses.
Reviews of such experiments, and connections to results from the large hadron collider can be found in Ref. \cite{Murata}. 
\\

%=== Data sequences ============
\begin{figure}[thb]
\includegraphics[width=7.0cm]{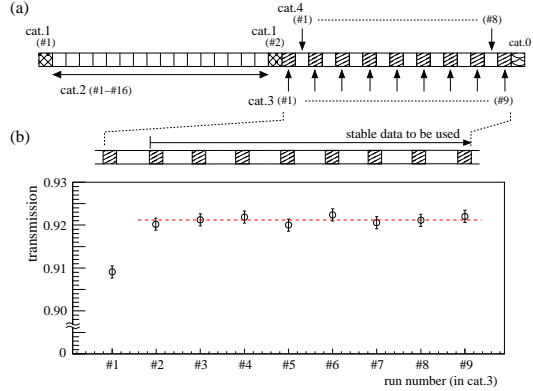}
\caption{\label{d_seq}
(a) Time sequence of taken data categories.
Category 2 contains scattering data taken when the chamber evacuated down to a level of $10^{-4}$ Pa,
representing measurements of the contribution due to the chamber, only.
Category 4 contains scattering data taken with the chamber filled up with xenon gas of around 167 kPa,
representing measurements of the contributions of the gas and the container.
(b) Time variation of neutron transmission through the 250 mm xenon gas, 
which is evaluated using the cat.3 data.
Data in stable condition of the period from \#2 to \#9 are used for the new-interaction search.
}
\end{figure}
%===============

\section{Setups and Data Sets}
The beam line consists of a neutron velocity selector, 
a neutron flux monitor, a collimation system, a scattering chamber which contains xenon atomic gas of 167 kPa pressure,
and a 2 dimensional neutron detector with 1 m $\times$ 1 m sensitive area.
We use an unpolarized neutron beam supplied from the ILL research reactor of 58 MW power.
A wavelength of 6 \AA\ was selected with 10\% FWHM fractional resolution.
Circular neutron apertures of diameters 20 mm was located  2.80 m and 0.22 m upstream from 
the center of the xenon-filled chamber, defining the beam divergence to be around 8 mrad.
The scattering chamber had a cylindrical shape  of length 250 mm and diameter 130 mm.
It was mounted on a leveling stage with a thermal insulating plate.
The cylinder axis was aligned along the beam traveling line using a laser light.
Outgas rate of the chamber had been measured before the beam time to be $0.5$ Pa/h,
which corresponds to contaminations of less than 75 ppm in our successive measurements.
The chamber was connected to a getter-based gas purifier, a vacuum system, and a high-pressure xenon gas bottle.
The vacuum system could evacuate the chamber volume down to at least $10^{-4}$ Pa level.
The main detector on the D22 beam line consists of 128 $^3$He filled tube counters arranged vertically and horizontally spaced by 8 mm.
The neutron position along the tube axis could be determined with 4 mm resolution.
Therefore the detector enabled us to measure the 2 dimensional scattering distribution image of 
8 mm (horizontal) times 4 mm (vertical) pixels. 
The detector was located $3005 \pm 3$ mm downstream from the center of the scattering chamber.

Data sets are categorized to five:
background data (cat.0),
data of transmission through the evacuated chamber (cat.1), 
data of scattering from the evacuated chamber (cat.2), 
data of transmission through the xenon gas filled chamber (cat.3), and
data of scattering from the xenon gas filled chamber (cat.4).
The transmission data are taken with an attenuated direct beam, 
where a beam stopper is moved out of the beam axis.
Total accumulation times are 20 minutes, 10 minutes, 16 hours, 50 minutes, and 24 hours for cat.0, 1, 2, 3, and 4, respectively.
The cat.3 and cat.4 data are taken alternately to monitor the stability of xenon gas pressure.
Time sequence of the taken data is illustrated in the Fig \ref{d_seq}(a).
Figure\,\ref{d_seq}(b) shows a variation of the neutron transmission through the xenon gas.
The gas pressure varied during the first period of cat.4 (\#1),
so we use the data from cat.4 (\#2) to (\#8) for the new-force evaluations.
The transmission is measured to be $0.921 \pm 0.001$ during the stable period.

%=== Measurements/Limits =====
\begin{figure}[thb]
\includegraphics[width=11.0cm]{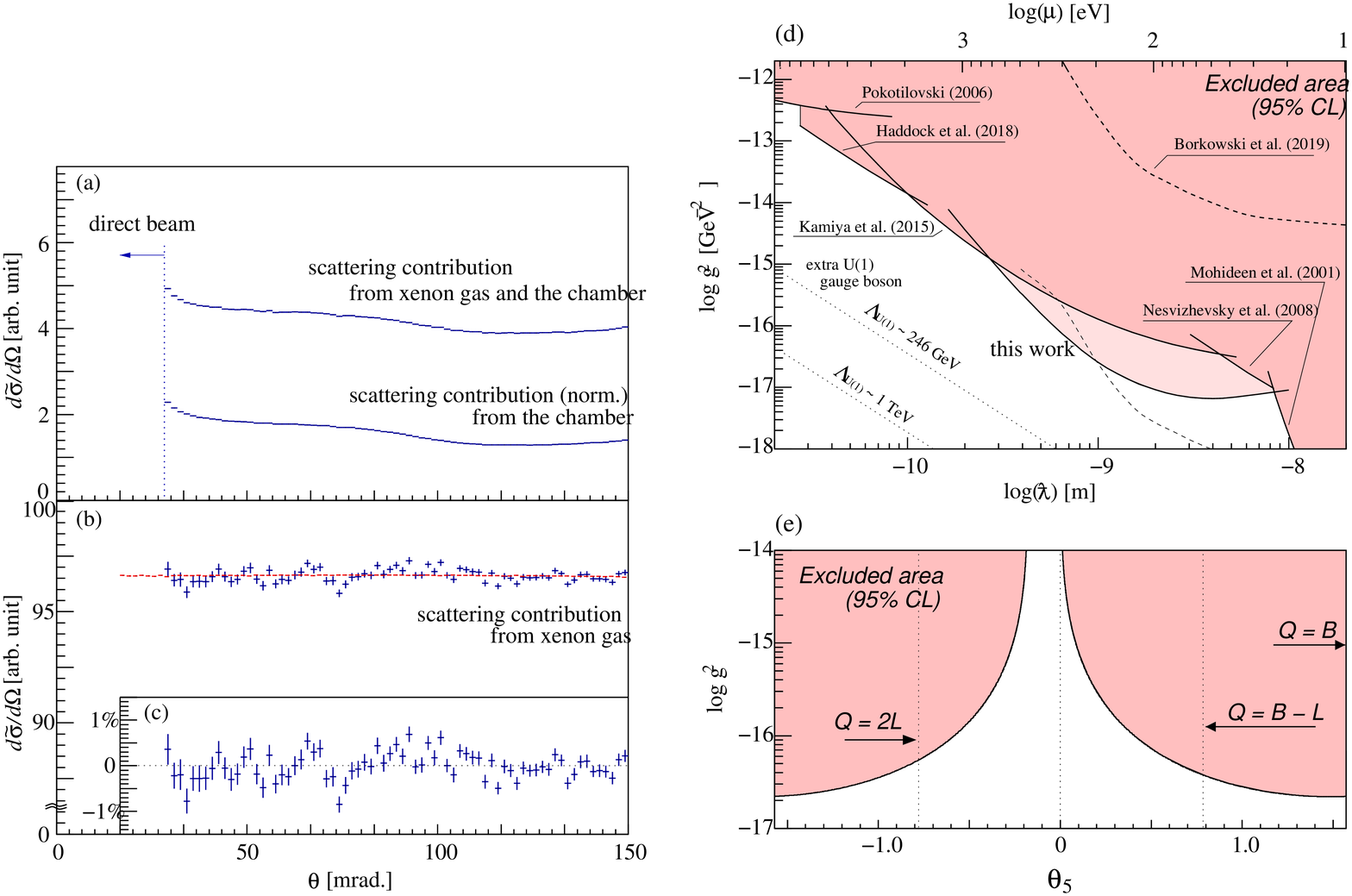}%
\caption{\label{meas}
(a)
Measured angular distributions for the cat.4 data (above) and for the cat.2 data (below).
The cat.2 distribution is normalized using the transmission of xenon gas.
The contribution from beam background of the cat.0 data, not shown in the figure,
is a flat distribution and smaller than the scattering data by three orders of magnitude.
(b)
Extracted angular distribution due to the scattering off the xenon gas.
The red dashed line shows a best fit (see text).
(c) Residuals from the best fit function.
(d) Obtained 95\% CL limits in the Yukawa-type $g^2$ - $\lambdabar$ parameter space ($\lambdabar \equiv 1/\mu$).
The upper dashed line represents 
a result using spectroscopy of weakly bound Yb$_2$ molecule \cite{Borkowski:2019}.
The lower dashed line shows the prospective precision of this new method.
Theoretical predictions due to extra U(1) gauge bosons are shown as dashed lines for
symmetry braking scales of $\Lambda_{U(1)} \sim 246$ GeV and $ \sim 1$ TeV \cite{Fayet:1993fu}.
(e) Re-interpretation of our new limits for $\lambdabar = 1$ nm (as an example) to the model in which a coupling charge $Q$ is expressed 
as a linear combination of the baryon and the lepton number, $B$ and $L$, as a function of the mixing angle $\theta_5$.
}
\end{figure}
%===============

\section{Measurements and New Limits}
Figure\,\ref{meas}(a) shows the measured scattering contributions with and without xenon gas.
They have been normalized by the transmission and integrated counts on the neutron flux monitor for each data category.
The difference between the two distributions, which is reflected by the $q$-dependence of the neutron-xenon scattering processes,
is shown in Fig. \ref{meas}(b) with the best fit curve of the fitting function 
$ {\rm d}\tilde{\sigma}/{\rm d}\Omega (\theta; \beta) = (1 - \alpha^{*})(1 - \beta) h_{c}(\theta) + \alpha^{*}(1 - \beta) h_{em}(\theta) + \beta h_{new} (\theta; \mu)$
for $1/\mu = $ 1 nm.
Here, ${\rm d}\tilde{\sigma}/{\rm d}\Omega (\theta; \beta)$ represents 
a smoothened differential cross-section evaluated by convoluting the Eq. (2) with the finite beam size,
the length of the scattering target, neutron attenuation in xenon gas, 
and the thermal motion of xenon gas atoms at 293 K using the Monte Carlo method.
The quantities $h_{c}(\theta)$, $h_{em}(\theta)$, and $h_{new} (\theta; \mu)$ are convoluted distributions corresponding to
the first, second, and the third term in the equation\,(\ref{eq:cs}), respectively.
The Monte Carlo estimation assumes isotropic thermal motion of Maxwellian distributed atoms.
This thermal motion leads to a slight forward-backward asymmetry 
and hence inclination of the expected distribution.
This effect mimics an additional $q$-dependence in the scattering processes,
which becomes notable for lighter targets, such as for argon gas.
The measured distribution is normalized to $\int_{S} d\tilde{\sigma}/d\Omega(\theta) d\theta = 1$
over the signal acceptance $S$ of 28 mrad $ < \theta < $ 150 mrad,
providing a high sensitivity to ranges of a new interaction around the nanometer scale.
In the fitting function, $\alpha^{*}$ is a constant calculated from measurements of the neutron mean-square charge radius \cite{Tanabashi:2018oca},
and $\beta$ is a fitting parameter, which represents the strength of new forces $g^2$.
Residuals from the best fit curve are shown in the Fig.\,\ref{meas}(c).
The best fit values of $\beta$ are, for example, $(-1.7 \pm 1.6) \times 10^{-3}$ and $(-9.6 \pm 5.6) \times 10^{-4}$ 
for new forces of 0.4 nm and of 1 nm range, respectively.
Using the Feldman and Cousins interpretation for the estimation around the physics boundary \cite{Feldman:342419},
we set improved limits for new gravity-like interactions as shown in the Fig. \ref{meas}(d).
Estimations of systematic effects on beta
are dominated by the uncertainties of the neutron transmission $\gamma$ and the sample-to-detector distance.
A shift by the one-sigma uncertainties of these quantities would shift 
the best fit value of beta for new forces of 1 nm range by around 4\% and 6\%, respectively.
Other effects due to a non-Gaussian component of neutron wavelength distribution,
temperature of the xenon gas, calibration factors for each pixel,
and uncertainty of the atomic form factor
are estimated to be much smaller.

The obtained limits are re-interpreted within a model in which a charge of new interaction 
is expressed as a linear combination of the baryon number $B$ and the lepton number $L$.
This assumption was discussed in references \cite{Fischbach:1988jv, NonAcce:1995}
and experimental constraints in a hundreds micron range on this picture are reported in \cite{Adelberger:2007et}.
Here we follow the notation of the charge of new interactions as in \cite{NonAcce:1995},
$Q = B \sin\theta_5 + (B-2L) \cos\theta_5~,$
where the mixing angle $\theta_5$ expresses a preference of the new force
to couple to baryons or leptons.
The leptophilic forces would imply $\theta_5 = -\pi / 4$.

\section{Conclusion}

We measured a neutron-xenon scattering distribution and obtained improved limits for gravity-like non-Newtonian forces in the nanometer scale. The results were re-interpreted with a model with new charges which are expressed as a linear combination of the baryon and the lepton number.

\begin{acknowledgments}
We appreciate Mark Jacques for technical supports on the instrumentation.
This work is supported by JSPS KAKENHI Grant No. 17H05397, 18H01226, and 18H04343.
\end{acknowledgments}

\nocite{*}
\bibliography{D22_APPC19}% Produces the bibliography via BibTeX.

\end{document}